# JTrack: A Digital Biomarker Platform for Remote Monitoring in Neurological and Psychiatric Diseases

**Authors**

Mehran Sahandi Far[1,2], Michael Stolz[1], Jona M. Fischer[1], Simon B. Eickhoff [1,2], Juergen Dukart[1,2][*]

[1] *Institute of Neuroscience and Medicine, Brain & Behaviour (INM-7), Research Centre Jülich, Jülich, Germany*
[2] *Institute of Systems Neuroscience, Medical Faculty, Heinrich Heine University Düsseldorf, Düsseldorf, Germany*

**\*Corresponding authors:**
**Juergen Dukart, PhD**
Institute of Neuroscience and Medicine, Brain & Behaviour (INM-7), Research Centre Jülich, Jülich, Germany
Email: juergen.dukart@gmail.com
Phone: +49 2461 61 3187

**Abstract**

**Objective:**

Health-related data being collected by smartphones offer a promising complementary approach to in-clinic assessments. Here we introduce the JTrack platform as a secure, reliable and extendable open-source solution for remote monitoring in daily-life and digital phenotyping.

**Method:**

JTrack consists of an Android-based smartphone application and a web-based project management dashboard. A wide range of anonymized measurements from motion-sensors, social and physical activities and geolocation information can be collected in either active or passive modes. The dashboard also provides management tools to monitor and manage data collection across studies. To facilitate scaling, reproducibility, data management and sharing we integrated DataLad as a data management infrastructure. JTrack was developed to comply with security, privacy and the General Data Protection Regulation (GDPR) requirements.

**Results:**

JTrack is an open-source (released under open-source Apache 2.0 licenses) platform for remote assessment of digital biomarkers (DB) in neurological, psychiatric and other indications. The main components of the JTrack platform and examples of data being collected using JTrack are presented here.

**Conclusion:**

Smartphone-based Digital Biomarker data may provide valuable insight into daily life behaviour in health and disease. JTrack provides an easy and reliable open-source solution for collection of such data.

**Introduction**

Neurological and psychiatric diseases typically present with symptoms that are complex, atypical, fluctuant in disease progression, and display high variability between patients [1]. Current diagnostic and efficacy evaluation methods often rely on in-clinic visits and subjective evaluation by patients, care-givers or clinicians. In-clinic evaluation methods are often costly, time-consuming and limited in their quality and quantity of observations [2]. In addition, they are often prone to high inter- and intra-rater variability [3]. The aforementioned drawbacks of traditional diagnosis methods may affect the diagnostic process especially in the early stage of the disease where there is a lag between the onset of the pathological process and the onset of symptoms [4].

Psychiatric and neurological diseases are typically long-term illnesses that cause significant fluctuations in symptoms over time. Therefore, recall and reporting biases are the key difficulties in evaluating respective diseases in episodic in-clinic visits. Remote monitoring of patients in their everyday-life using sensor-based at smart technologies is rapidly evolving and may assist clinicians in facilitating early diagnosis and evaluating and adjusting interventions. There has been an evolving interest in using newly emerged smart sensor technologies for monitoring of patients [5–10].

Modern smartphones and wearables are equipped with various sensors including motion (i.e., acceleration, gyroscope), location (i.e. Global Positioning System (GPS)), environment (i.e. barometer, temperature, light) and health sensors (i.e. heart rate). This rich combination of sensors along with their ability to collect ecological momentary assessments (EMA), and information about social interaction (i.e., social media, messaging and phone calls) have made smartphones a potential alternative to in-clinic evaluation for various types of assessments. Such health-related information being collected in clinical trials are often referred to as digital-biomarkers (DB) [11]. DBs can provide objective, ecologically valid, and invaluable information for better understanding of specific diseases. In addition, DBs enable frequent assessments from larger target populations over longer periods of time and may thus provide detailed insight into inter- and intra-individual disease variability in daily life.

Several contributions enabling the use of smartphones as an assessment tool have been recently introduced. The first set are commercial devices such as Fitbit, Garmin, Apple and Samsung devices [12–15]. The main focus of these applications is to provide feedback on the daily activity of users by visualizing and showing notification regarding their heart rate, number of steps and kind of activity. However, most of these devices provide limited access to the raw data and do not support high-frequency data collection. A second type are applications and platforms developed by researchers such as AWARE [16], RADAR-base [17], Beiwe [18], mCerebrum [19], mPower[20] and many others. The main focus of these mostly open-source platforms is to enable data collection for research applications as well as to facilitate data sharing and reproducibility. Yet, these software packages are often limited by an often narrow focus to some specific clinical indications or with respect to privacy aspects. Also, these once in a while updated platforms make some of them unstable for the rapidly growing smartphone ecosystem. While there are several platforms able to collect context-driven data, the trade-off between privacy, optimization, stability and research-grade data quality is not well met, yet. Thus, we aim to fill this gap by introducing the JTrack platform.

JTrack platform is a solution for clinicians and researchers to collect, manage, and share digital biomarker data. JTrack was developed as an Android-based application for smartphones and an online server-side dashboard. JTrack application comprises the following main categories of components: sensor data, location data, Human Activity Recognition (HAR), and smartphone and application usage monitoring. Each component has the option to be used for active (with user interaction) and passive (without user interaction) monitoring. The dashboard side is an online platform to create and manage studies, integrating DataLad [21] infrastructures to facilitate management and sharing of collected data.

**Methods**

[Figure 1]
[Figure 2]
[Figure 3]

**General description**

Here we introduce the main components of the JTrack platform (Figure 1) comprised of the JTrack app (Figure 2) and an online dashboard interface (Figure 3).

The smartphone application JTrack was developed for smartphones with the Android operation system (OS). The reason for selecting Android was a wider range of users (73% [22]) and fewer restrictions which were necessary for technical aspects of application development.

JTrack enables passive 7/24 data collection running in the background. Active data collection is enabled through simple interaction (i.e. start and stop recording, i.e. before and after execution of a specific task). All collected data are recorded locally in the application and then synchronized on a periodic basis (i.e., connection to the Internet, have enough battery charge). All local data are deleted from the phone storage upon successful data transfer. To minimize the risk of data loss, we implemented auto-start functionality (without user interaction) to resist unwanted application crashes or operating system reboots, and all the crashes are reported via the Firebase dashboard [23].

On the server side, the JTrack dashboard was developed as an online web-application where study owners can create and manage studies. The dashboard consists of a front-end interface and a back-end API which is integrated with DataLad [21] as a data-management tool. The dashboard provides an overview of received data including sanity checks such as MD5 for received data, and embedded validity checking methods.

**QR-Code Authentication**

To provide a convenient and secure way of activation we implemented a QR-code method. The QR-code for each subject is generated as a pdf file from the dashboard. Each QR-code contains all the necessary information such as user ID, Study ID, and address of the target server or an optional authentication method (e.g. OAuth2). To join a study, the one-time QR-code needs to be scanned using the QR-scanner embedded in the JTrack app. Additional backup QR-codes are provided for scenarios in which users may want to leave and re-join or need to switch their device.

**Location service**

Location service provides an update on visited location data such as longitude, latitude, and attitude. This service operates as a part of the passive recording. The location data can be inquired based on pre-defined periods (i.e. 10 minutes). To ensure anonymization, for each user, a random value is generated during activation on the phone, which shifts the latitude to a random place on the globe. This value remains on the phone throughout the study and is deleted upon deinstallation of the app. To keep a high accuracy, each data point is first transferred from WGS-84 to Cartesian coordinates. After the transformation using the generated value the coordinates are transferred back to their native space. Since this transformation occurs before actual recording, all the collected data is relative and cannot be used to recover the user's actual location. Furthermore, we used a fused method that provides more accurate data (median accuracy of 14 meters) by combining GPS and network information.

**Human Activation Recognition (HAR)**

Inertial Measurement Unit (IMU) sensors of smartphones or wearables can be used to differentiate between human activities. Several studies described reliable algorithms for HAR [24,25]. Nowadays, these algorithms are routinely deployed in commercial devices, as well as in a wide range of research areas from medicine to military. JTrack uses the Google Play Activity Recognition Service [26] for HAR, which recognizes up to six types of activity (walking, running, still, on bicycle, on vehicle or tilting). JTrack can record the detected activity and the assigned certainty with a pre-specified frequency of five minutes. The HAR module is computationally lightweight, optimized and does not require direct access to raw sensor data.

**Application Usage Statistic**

JTrack can collect the statistic of user's interactions with the smartphone. This data includes the name of the application and the amount of time it is used in the foreground since the previous midnight. Phone calls and SMS are treated as applications with same usage statistics being collected as above. No content of the applications, messages or phone calls (including phone numbers) is collected at any stage.

**Sensors**

Various sensors are embedded in any modern smartphones which are classified as hardware implementation (i.e., accelerometer, gyroscope, barometer) or software implementation (i.e., rotation sensor). JTrack enables collection of data from most of the available sensors depending on the device model and version of Android. Among these, accelerometer, gyroscope and gravity sensors are the most important sensors for researchers focusing on motion analysis [6,27–31]. As a default, JTrack provides recording of accelerometer and gyroscope data in the passive collection mode. Others sensors can be added upon the researcher's choice by using the provided template module which requires minimal coding effort. For each sensor, sampling frequency in Hz can be adjusted using the dashboard when creating a new study.

**Dashboard**

When creating a new study in the dashboard, all aspects of a study such as study name, duration, number of subjects, recording frequency, and categories of data to be collected can be customized. After creating a study, the dashboard will generate QR-codes which are used for enrolment into the study. The dashboard also provides management tools on an ongoing study producing information such as a number and time of received data for each sensor/modality and status (i.e., active, not active) of each participant in a particular study. We also implemented several quality controls including highlighting of missing data.

Furthermore, to assist study managers to establish further interaction with participants, we embedded a messaging method in the dashboard which allows to send a push notification directly to the participants' phone, either by selecting specific subject numbers or all participants within a particular study. Layered design (backend, frontend and data management layer) also makes the Dashboard flexible and extendable for further interaction and integration with third-party applications.

**Performance, Security and Privacy**

At all stages of the development, attention was paid to security and privacy as a main priority. In this context, we designed the JTrack platform to comply with GDPR and Google Developer Policies [32]. No sensitive data such as name, phone number, phone contacts or actual location are recorded at any stage. All the collected data transferred via Hypertext Transfer Protocol Secure (HTTPS) protocol and checked for any inconsistency using MD5 sanity checksum.

Concerning patient privacy, all users using JTrack are provided with clear information on what is been recorded and why. Permission requests for each module need to be approved during installation and activation. All participants may also stop and leave a study at any time directly from the app. Also, remote configuration and one-step recording allow clinicians to gain optimum control over the collected data without the need to collect any identifying information.

To reduce battery and memory usage, we provided several built-in optimizations such as:
- Detecting still period of the phone to pause recordings
- Delete locally stored data right after synchronization with the server

- Scheduled synchronization based on predefined criteria such as access to a Wi-Fi connection
- Detect and provide a possibility to bypass performance optimizations (i.e., battery and memory) policies of phone manufacturers introduced on-top of the Android OS.

To reduce data loss due to crashes or reboots, automatic re-starting is implemented alongside with Firebase integration to obtain performance and crash reports.

**Results**

[Figure 4]

To illustrate the utility of the JTrack application sample data were collected in the beta testing phase. Figure 4 displays such sample data collected for each modality.

[Table 1]

**Comparison to other open-source digital biomarker solutions**

Table 1 shows how the JTrack platform compares to other similar and related platforms in terms of some key features such as security and privacy, activation, management and also stability.

AWARE [16] is a platform for remote assessment of a wide range of phone sensors, activity and self-reported data. AWARE also supports additional plugins for external sensors and new data. However, this ability also requires a further declaration of permission which limits control over privacy.

mCerebrum [19] is another platform for remote assessment supporting a wide range of high-frequency sensors with a focus on energy-optimization. However, this platform has not been updated in while (latest update is May 2018 in their GitHub repository), questioning its performance on new versions of Android-OS.

Beiwe [18] is the next platform supporting remote monitoring and DBs assessments which has a flexible study portal, modelling and data analysis tools. Nevertheless, this platform does not have local data storage and makes use of Amazon Web Services (AWS) cloud computing infrastructure which may limit the control of collected data and require further steps for deployment. Another drawback of this platform is the collection of identifiable data such as phone number, media access control (MAC) address of WIFI and Bluetooth devices.

RADAR-base [17] is the last open-source platform in the list. It has a well-organized structure which is using Confluent and Apache Kafka services and flexible study portal. Nevertheless, the deployment and adaptation of this platform require heavy configuration. Concerning the convenient registration, it requires text-based registration.

Location data being collected in background is considered as a big concern in terms of privacy which is also frequently regulated by Google Developer Policies [32] and restricted by recent updates in Android OS. Among all the compared platforms only RADAR-base provides relative location. Easy one-step registration and authentication via QR-Code, as well as remote configuration, make JTrack more practical in both the usage and management aspects. Battery and memory optimizations offered by Android OS or phone manufacturers can affect the stability and consistency of data collected, JTrack provides built-in detection and circumvention methods for better stability that did not achieve on comparator platforms at this level.

**Discussion**

We developed JTrack as an open-source, smartphone-based platform for digital phenotyping. JTrack consists of a smartphone application and an online dashboard enabling remote data collection and study management.

From the functionality perspective, many of the existing solutions were developed with the focus on specific applications, i.e. a specific disease (i.e. RADAR-base[17] and Beiwe [18]). Their application is therefore limited to a specific context. In contrast, some other platforms were developed to collect as much information as possible with little attention to data privacy (i.e. AWARE [16]). Such frameworks violate GDPR and Google Play Store policies limiting their deployment for many clinical applications. JTrack aims to fill this gap by providing a customizable platform that can be deployed across different indications whilst paying large attention to privacy and security policies. JTrack aims to comply with GDPR regulations as well as with the Google Play Store policies. It only requires minimal access to the device information and avoids collection of identifiable or sensitive data.

Developing an application for smartphones always requires dealing with variation in devices (e.g. manufacture, screen size, available sensors) as well as the variation of operation systems (OS) versions. Different manufacturers may add further OS optimizations such as limiting background processes. This may cause inconsistencies in performance of monitoring applications. We introduced several layers to detect, report and prevent the side effects of these variations. JTrack is actively maintained and covers up to 84.9 % of Android smartphones (Minimum Software Development Kit (SDK) 23) dealing currently with Android optimizations from eight main Android smartphone manufactures.

Potential applications for JTrack include but are not limited to monitoring of motion information in diseases associated with alterations of gait and other motor functions affecting phone use. Similarly, the ability to track phone usage allows for monitoring of different types of behaviour, i.e. phone-based social interaction. As such, JTrack may be useful to track such behaviours in healthy participants as well its alterations by specific disorders.

Finally, to facilitate the reusability, JTrack is released under open-source Apache 2.0 licenses. All modules including online-management dashboard can be adopted and extended. It has been designed with modular structure to enable flexibility and customization to support new data and sensors.

**Limitation**

Variation in device model, Android version, network quality, and other technical features may have negative effects on the performance of JTrack. Despite the effort to minimize crashes and data loss, there is no guarantee for such. During the development process, we used different third-party services (e.g. Google Play Service), any change or deprecation in these services, or Android policies may also affect the functionalities of JTrack partly or as a whole. Lastly, JTrack was designed and tested for smartphones. It may be used on other devices such as wearables (i.e. smartwatches) or tablets but further tests should be considered beforehand.

**Future works**

JTrack is an active and open source project which is continuously maintained. We consistently improve and add new features to the platform. The features described here are part of the v1 release. Newer versions may differ at the time this article is published.

**Data availability**

For the most updated and previous versions please visit the public repository accessible at https://github.com/Biomarker-Development-at-INM7

**Acknowledgements**

This study was supported by the Human Brain Project, funded from the European Union's Horizon 2020 Framework Programme for Research and Innovation under the Specific Grant Agreement No. 785907 (Human Brain Project SGA2).

**Author contribution**

MSF wrote the android based application and performed all analysis. MSF wrote the manuscript with support of JD. MS and JF wrote the dashboard online application. JD designed the overall study. SBE contributed to the overall design of the study. All authors reviewed and commented on the manuscript.

**Competing interest**

JD is a former employee and received consultancy fees on another topic from F. Hoffmann-La Roche AG. All authors report no conflicts of interest with respect to the work presented in this study.

**Table 1.** Comparison of existing frameworks with JTrack.

| Framework | Location Anonymization | Official app Stores | Data Management | Remote Configuration | Activation | Customized OS Detection |
|---|---|---|---|---|---|---|
| AWARE [16] | NO | NO | SQL | YES | Text-based | NO |
| RADAR-base [17] | YES | YES | MongoDB | YES | QR-Code | NO |
| Beiwe [18] | NO | YES | PostgreSQL | NO | Text-based | NO |
| mCerebrum [19] | NO | NO | MySQL | NO | Text-based | NO |
| JTrack (this study) | YES | YES | DataLad | YES | QR-Code | YES |

**Figure 1.** JTrack Platform's overview.

**Figure 2.** JTrack Application Environment. **a)** opening page of the application, **b)** requesting for camera permission, **c)** QR-Code scanner, **d)** requesting for location permission, **e)** requesting for an activity detection permission, **f)** referring to ask for usage permission which is used for application usage module, **g)** detecting a custom optimization and asking for disabling it, **h)** the main menu of the application, where users may access the administration menu, leave a study, get information about the application and do manual synchronization, **i)** administration menu, here we have access to the main setting of application, the information provided here is for further administration from study owners and most of the information is catch from sever during Installation.

**Figure 3.** JTrack Dashboard Environment. **a)** Main menu, **b)** here a new study can be created by specifying its details such as duration of the study, number of users, and list of data categories to be collected, **c)** Currently ongoing studies and details of the selected study can be found here. Also, the generated QR-Code can be downloaded here, **d)** To accomplish more interaction with users participated in a particular study, a message as a push notification can be sent to a target user(s), **e)** Details of received data, date of registration, date left, duration within in study and quality control by colour-labelling for sent data for users in a selected study can be monitored here.

**Figure 2**. Sample data from different modules of JTrack. **a)** Travelled distance and relative geolocation information for different days, **b)** distribution of different physical activates during a different time of different days, **c)** type of activity data combined with geolocation information, **d)** amount of time spent in different application types during one day, **e)** raw data recorded from the acceleration sensor, **f)** raw data recorded from the Gyroscope sensor.

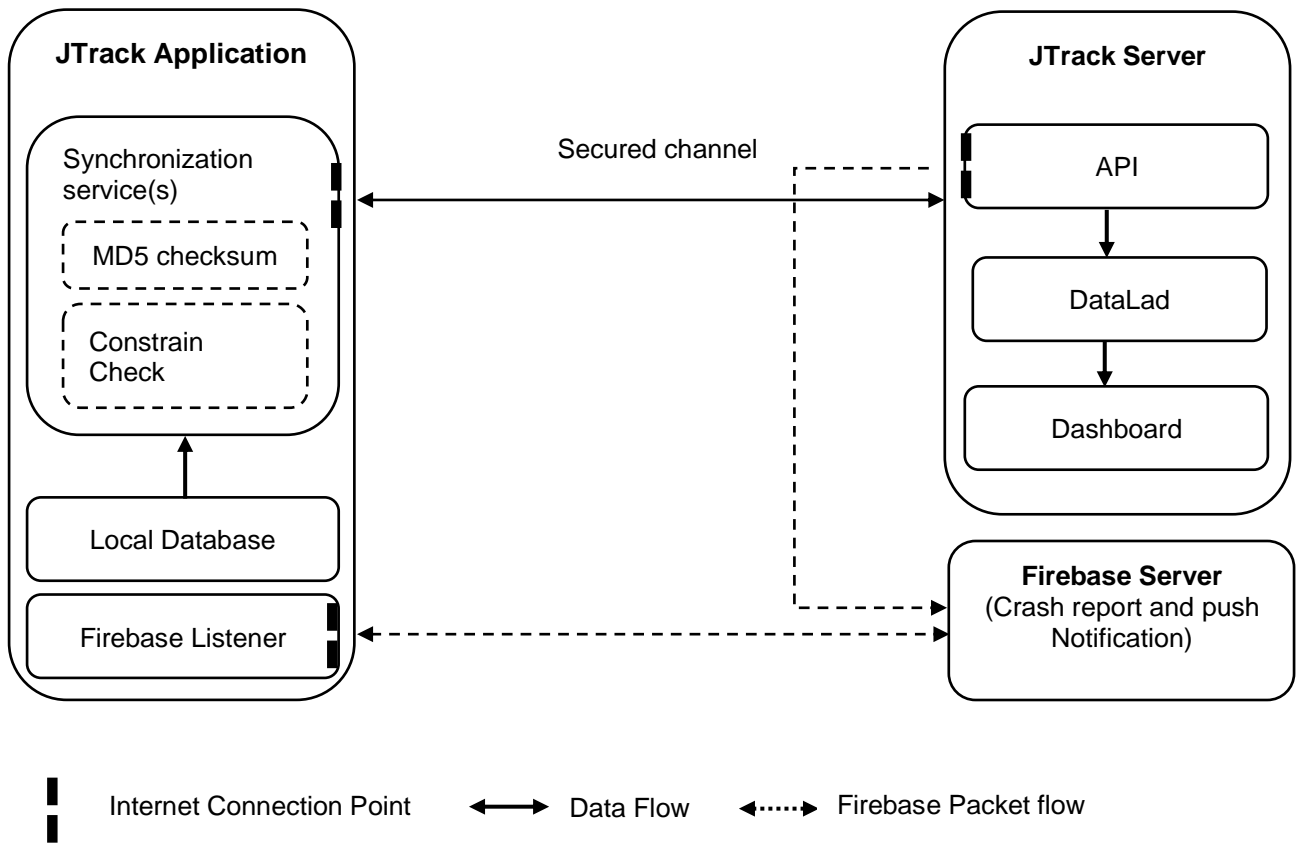

[Figure 1]

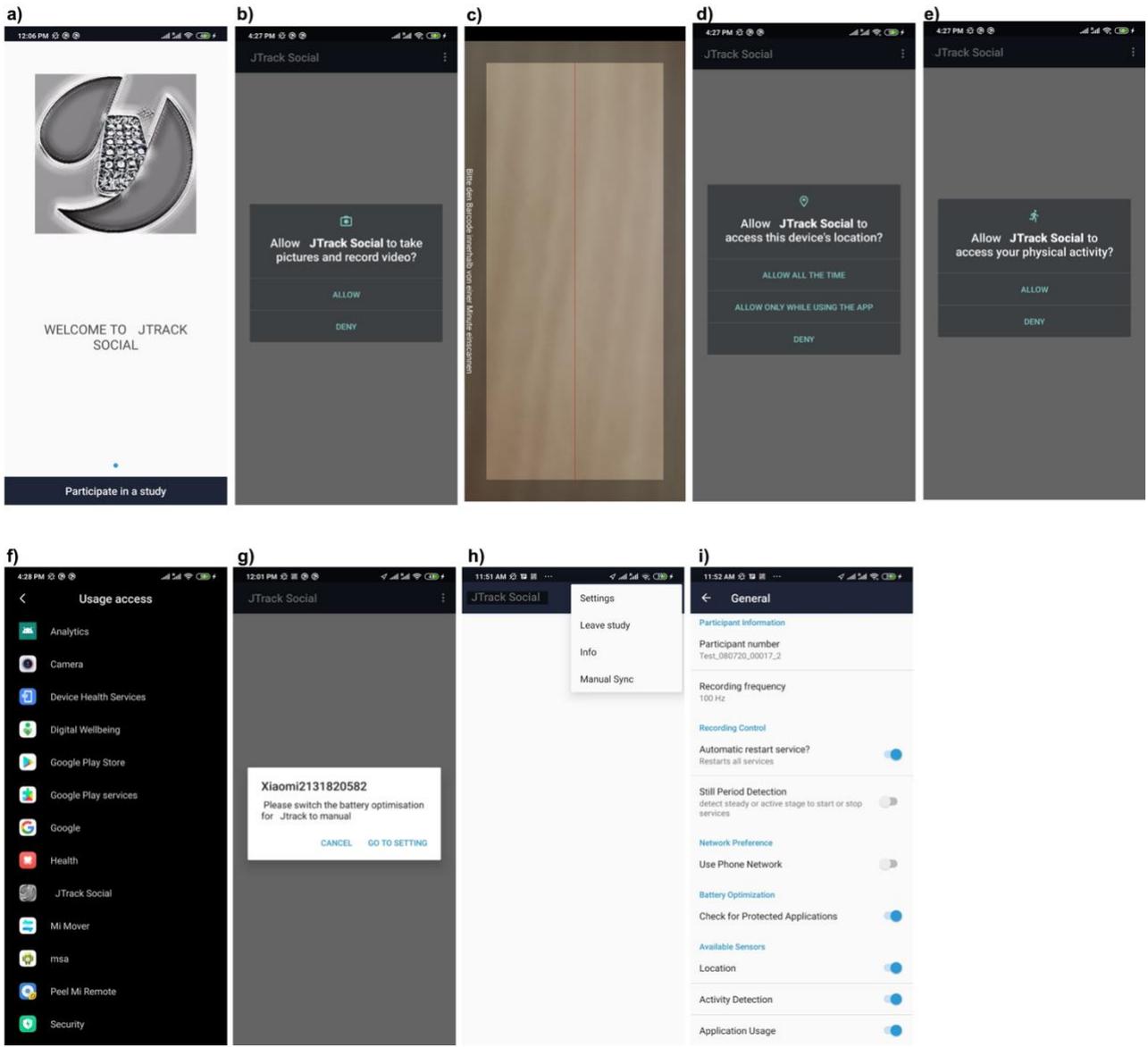

[Figure 2]

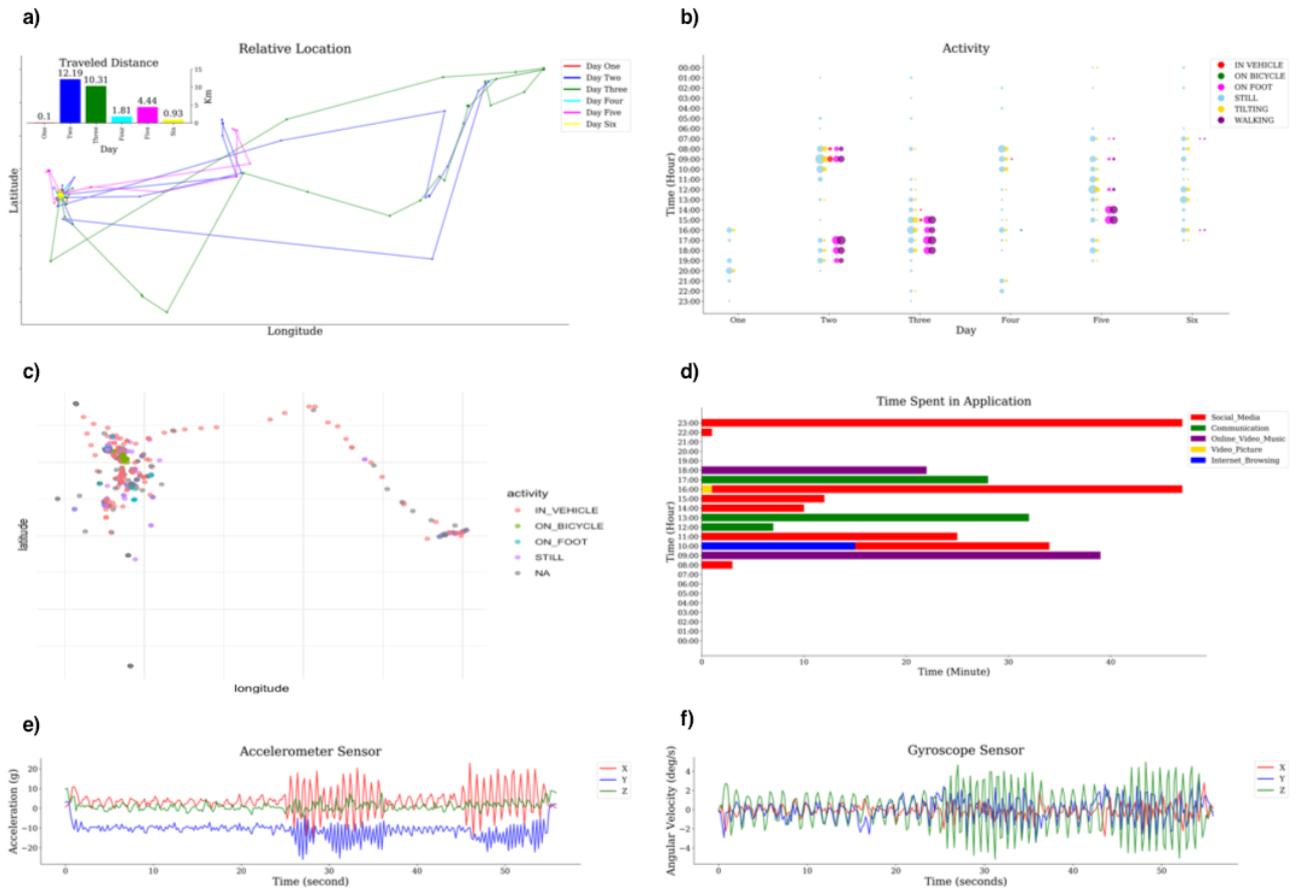

[Figure 3]

[Figure 4]